\documentclass{revtex4}
\usepackage{makeidx}  

\usepackage{graphicx}

\begin{document}
%
%
\title{QKD in Standard Optical Telecommunications Networks}
%

%
\author{D. Lancho$^1$, J. Martinez$^1$, D. Elkouss$^1$, M. Soto$^2$ and V. Martin$^1$}
\address{$^1$ Facultad de Inform\'atica, Universidad Polit\'ecnica de Madrid,\\
Campus de Montegancedo, Madrid 28660, Spain}
\email{Vicente@fi.upm.es}
\address{$^2$ Depto. Seguridad en Redes y Servicios,
Telef\'onica Investigaci\'on y Desarrollo,\\
Emilio Vargas 6, Madrid 28043, Spain.}


%


\date{08 January 2010}

\begin{abstract}        
To perform Quantum Key Distribution, the mastering of the extremely weak signals carried by the quantum channel is required. Transporting these signals without disturbance is customarily done by isolating the quantum channel from any noise sources using a dedicated physical channel. However, to really profit from this technology, a full integration with conventional network technologies would be highly desirable. Trying to use single photon signals with others that carry an average power many orders of magnitude bigger while sharing as much infrastructure with a conventional network as possible brings obvious problems. The purpose of the present paper is to report our efforts in researching the limits of the integration of QKD in modern optical networks scenarios. We have built a full metropolitan area network testbed comprising a backbone and an access network. The emphasis is put in using as much as possible the same industrial grade technology that is actually used in already installed networks, in order to understand the throughput, limits and cost of deploying QKD in a real network.

\end{abstract}
\maketitle              

\section{Introduction and Testbeds}

  To date, QKD development has mostly focused in non-shared point-to-point links and in networks 
made of these links~\cite{secoqc}. However, the benefits of  integrating QKD with the commercial network infrastructure are clear. With the poss\-i\-bi\-li\-ty of a scalable deployment and without requiring a massive initial investment, QKD can reach a broader market. Current networks are evolving towards  optical, passive infrastructures and this opens up a window of opportunity for QKD integration, since an all optical path among two points in the network is no longer unfeasible nor extremely expensive. However, the simultaneous propagation of quantum and classical signals over a shared link, presents important problems due to the spilling of photons coming from classical signals that typically have 100 dBm more power.

Research has been done about the use of QKD in optical networks~\cite{toliver03,runser05,xia05}, including some recent studies~\cite{toliver09} using the 200 GHz (1.6 nm) ITU DWDM grid (Dense Wavelength Division Multiplexing). Their aim was to characterize the sources of noise and to demonstrate the use of reconfigurable optical add and drop multiplexers (ROADM) for QKD purposes. Here we go one step beyond in completing a metropolitan area optical network able to establish a transparent path among two of their nodes in order to sustain a QKD link.

Metro networks are logically divided into backbone and access parts. The backbone is specialized in high speed communications. The number of nodes is limited and the cost per node is less of a concern, sharing importance with throughput, reliability, serviceability and upgradeability. Clients are connected to the backbone through an access network. This is a point to multipoint network, with one end  connected to the backbone at the 
carrier company premises, while the other gives service to several clients. A shared link goes from the backbone to some form of splitter that  is located in the vicinity of the clients. An exclusive use link connects the splitter to each one of the clients. In the access network, throughput is not as relevant, but cost and maintenance are important issues, since many of them have to be deployed and part of the equipment is either sold or leased to the client.

The new generations of access networks, that are being massively deployed in the current fiber to the home infrastructures, are designed as optical and passive, which opens the possibility of using them for QKD.  

The purpose of the present paper is to investigate the limits, trade offs and compatibility issues that the integration of QKD with the existing or newly deployed optical infrastructure could pose.  In order to do so, a full metro network comprising backbone and access parts is used.  

The backbone testbed is built as a ring composed of three CWDM (Coarse WDM~\cite{ITU_CWDM}) ROADM nodes, the most common technology in metro and regional networks. Typically,
the CWDM ITU grid is implemented in a spectrum of 18 channels separated 20 nm starting at a central wavelength of 1270 nm. The bigger channel spacing makes for an easier filtering of the spurious photons 
coming from other channels and made the use of amplifiers less 
common in this kind of networks, since no single EDFA amplifier 
is able to amplify all of the CWDM channels, thus making the
option less attractive from a cost/benefit perspective than in DWDM links.

The access network uses the GPON (Gigabit Passive Optical Network) standard~\cite{GPONStandard}, again, the most used in new deployments of fiber to the business. GPON multiplexes three wavelengths over the same fiber to connect the OLT (Optical Line Termination) on the backbone side to the ONT (Optical Network Termination) on the
client side. A splitter located at some point in between the OLT
and the set of the ONTs ---typically much closer to the latter--- divides the signal among all the ONTs. The standard defines the 1490 nm wavelength to transport the downstream channel, 1310 nm the upstream channel and 1550 nm for analog 
video broadcast. The last one is rarely used for its intended
purpose and we are using it for the quantum channel. This is just a matter of convenience and  other channel  could have been chosen.

 A brief report of these setups, was given in~\cite{lancho08}. It is to be noted that the emphasis is put in what can be readily achieved with currently deployable technology, using as much off the shelf systems as possible and limiting  to a minimum the modifications done to what, in essence, is a standard network environment.

\section{Results}

  Data were obtained using id Quantique two way QKD systems, models 3000 and 3100, using BB84. Their maximum admissible loss budget in order to obtain at least a few secure bits per second is around 15 dB. The maximum throughput under ideal conditions (QBER=0), imposed by the detector deadtime, is 100 kbits/sec. In our case, mean photon 
number was set to simulate a decoy state protocol~\cite{decoyLo,decoyHwang,decoyWang,Wang05} with signal plus one decoy. The optimal mean number for our setup, 0.79, was calculated 
according to~\cite{Ma05}.  A full protocol stack, including LDPC (Low Density Parity Check) error correction with 1.05 efficiency~\cite{LDPCnuestros} and privacy amplification was used.

The backbone results were obtained using three wavelengths.  Two wavelengths, 1510 and 1470 nm are used for classical signals (co and count\-er\-propa\-ga\-ting, respectively) while 1550 is reserved for the quantum channel. This is not an specially favorable case for QKD, where the quantum channel should have been located at the shorter wavelength. Modules to add and drop the 1550 channel as desired are included at each node.  Extra filtering to further isolate the quantum channel was needed and standard DWDM 100 GHz (0.8 nm) filters were used. It is to be noted that, due to the spreading of CWDM and DWDM technology, very narrow and high quality filters are starting to be readily available for commercial use. 50  GHz (0.4 nm) filters are now reasonably common, hence we have extrapolated the data to this case. Losses in this scenario, without the fiber, are 8 dB. A full description will be published elsewhere.

\begin{figure}[!t]
\centering
\includegraphics{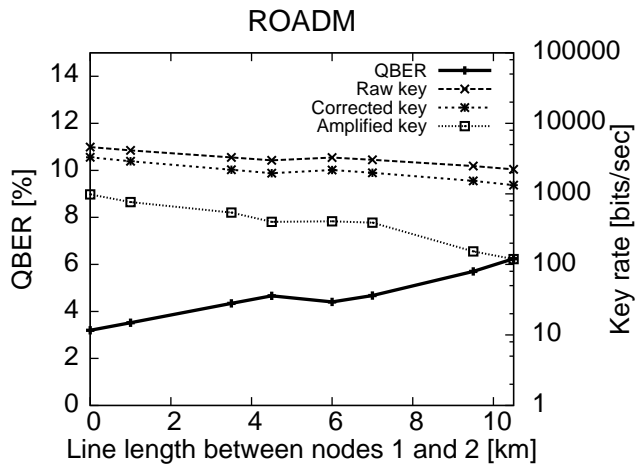}
\caption{ Results obtained in the backbone testbed. Quantum Bit Error Rate and key rate is presented as a function of the fiber length connecting ROADM nodes 1 and 2, the worst situation for QKD.  QBER is depicted on the left axis while key rate in their different distillation phases, till completing privacy amplification, is on the right axis.  Data, obtained with 100 GHz (0.8 nm) filters, are extrapolated to 50 GHz (0.4 nm) filters. Final secure  key rate approaching half a kbit/sec. can be obtained at 6 km and even after 10 km there is a net key throughput of around 100 bits/sec. The change in trend that can be seen in the data before and after 4.5 km is due to the use of different types of fiber (both standard and commercial) that produce different amounts of Raman scattering. Note also that the different segments of fiber are joined through connectors and, for the longer distances, four of them were
needed. This means that (in regard to losses but not to Raman scattering) the actual line length is of around 14 km. The average distance among ROADM nodes in real metro network is usually well below 10 km. }
\label{BackboneResults}
\end{figure}

  In the backbone  experiment, two QKD systems are used to test point to point key growing, key pass-through and key forwarding 
 through the ring while data signals are being simultaneously transmitted at full rate. In Fig.~\ref{BackboneResults} we present an example of a point to point key growing using a pass-through configuration in the middle node. The Quantum Bit Error Rate and key throughput, at different stages of the key distillation process, are shown as a function of the fiber length connecting the first and second ROADM nodes. Only a short, fixed length, fiber was used among the second and third. This is a worst case scenario for QKD since it produces more spurious photons  due to Raman Scattering than the situation in which, for the same total distance, the second ROADM node is located somewhere else in between.
The QBER value obtained, although high, is well below the 11\% threshold above which no secret 
key rate can be obtained in the usual setting with one way error correction and privacy amplification.
A secure key rate around half a kbit/sec. was obtained for distances till 6 km, although is reduced to 100 bits/sec. for 10 km, 
a length larger than the typical for ROADM nodes. It is to be noted that these scenarios are dominated by absorptions in 
the nodes and not in the fiber. Fiber length is relevant as long as it contributes to produce Raman scattering. 
It is important to put this result in perspective: a fully populated DWDM link, 
with all the 160 channels transmitting at 2.4 Gbit/sec. and secured with a 1 kbit/sec. key, will have to encrypt less than $2^{37}$ bits per 256 bit AES key. This is less than the known reasonable security limit of $2^{40}$ for DES~\cite{vanassche}, an assumed lower security encryption method. For standard links, key rates as low as a few tens of bits/sec. would suffice to have a much higher security 
level than what is standard today.

The number of spurious photons entering the quantum channel depends on the total power sent through the fiber, but also on the fiber itself and some other characteristics like the directivity and crosstalk of the components used to build the ROADM, which in our case is a standard model with no special modifications. In a CWDM set up, four wave mixing is not an issue and only Raman scattering is a real threat. Apart from the line length, the only controllable variable is the total launch power. Here it was attenuated (differently for each link length) respect to the standard power used, but well above the detection limit of the transponders, so no increased data error rate resulted in the classical channels working at full speed.

\begin{figure}[!t]
\centering
\includegraphics{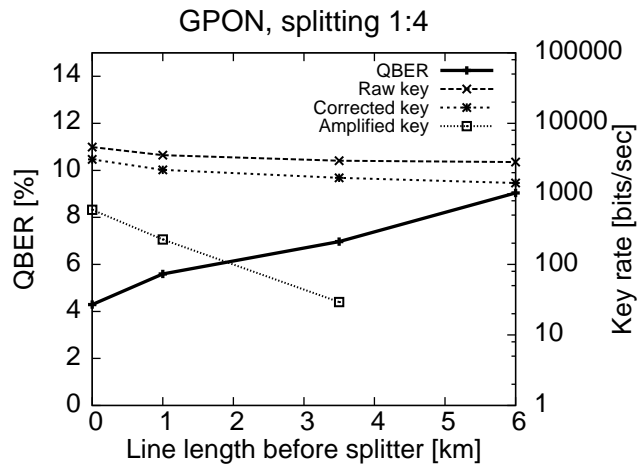}
\caption{ Results obtained in the access network. Quantum Bit Error Rate and key rate is presented as a 
function of fiber length connecting the OLT with the splitter,
the most populated part of the link and the worst configuration for QKD. Typical OLT-splitter distances can differ a lot with the geographical environment, but are usually on the shorter side for big cities.  QBER is depicted on the 
left axis while key rate, in their different distillation phases, till completing privacy amplification is on the right axis.  }
\label{GPONResults}
\end{figure}

 In the access network  experiment with GPON, a continuous data flux was established among the OLT (backbone side) and ONT (client 
side)  using the 1490 nm (downstream) and
1310 nm (upstream) channels. Again, QKD used the 1550 nm channel. Due to the more integrated nature of the access equipment (the client part needs to be compact and cheap) there are less modifications that can be readily done. The launch power is fixed and only a small attenuation can be introduced in the OLT and only in the 1490 channel. The filtering used was the same (100 GHz) and the maximum splitting factor possible with our QKD equipment is four. Losses without fiber are 9 dB. The results
are shown in Fig.~\ref{GPONResults}. This time, QBER and key rates are presented as a function of the line length connecting the OLT and the splitter, which is the longest segment in access networks and, again, the one that more heavily penalizes the quantum channel. This scenario is more demanding than the ROADM but this is mainly due to the difficulty of attenuating the total power in the line without HW modifications of the OLT/ONT. There is an advantage, however, in that  there is no actual need to have the fiber populated with classical signals at the same time that the quantum channel is being used. Because GPON works on a Time Division Multiplexing control, it is easy in theory to assign time slots to the quantum channel. In practice, this is more difficult, since it requires access to reprogram the micro controllers and that would require the involvement of the manufacturer.
 The QBER obtained is quite high, although still clearly below the threshold. It already starts, for a 0 km line, at 4\% due to the crosstalk in the internal components, mostly coming from the 1310 nm channel that is impossible to attenuate in our setup. This 
produces a secure key rate around 500 bits/sec. at 0 km that rapidly reduces to  20 bits/sec. at 3.5 km. In our next measurement, at 4.5 km, there is no secret key rate. This would call for a time slot assignment for the QKD channel or a redesign of the OLT/ONT pair to work on a much lower power budget.
Although the final key rate was lower than in the backbone, it is still enough to sustain a 256 bits AES  key renewal rate  faster  than it is usual today at short distances. It is to be noted that typical OLT/ONT distances are of the order of 1-2 km in big cities.

\section{Conclusion}

In this paper we have demonstrated the use of QKD within standard metropolitan optical networks comprising a 
backbone and access network with all the testbed  built around readily available systems and components. The emphasis is put in finding out the limits of QKD integration in real networks rather than in absolute security against any conceivable attack, since  finite key effects are not taken into consideration.  Some test cases are the worst possible for QKD and they are not the way in which QKD would be reasonably integrated in a network; they are just the most straightforward without resorting to modify the classical communications equipment. They serve the purpose of bounding the performance limits. Some freedom exists when integrating QKD with standard optical networks and a mixture of scenarios is what is to be expected in the real world. For example, in the backbone network, the existence of unused dark fiber for backup purposes is quite common, hence, in the ROADM scenario, bypassing heavily loaded nodes is a real possibility. In GPON, the mentioned time slot assignment for QKD is an obvious solution that is not difficult to implement and would not hit significantly the performance of the access network. A direct connection without shared access network is also a possibility for medium to big companies. In metro scenarios, where limits to the integration are imposed by noise and absorptions in the optical equipment rather than in the fiber, moving the quantum channel to the second window ($\sim$1300 nm) would be advantageous. All this considerations support the view that the integration of QKD in modern optical networks is a real alternative to exclusive use quantum links and trusted-nodes QKD networks. 

In the frame of the present testbed and with current QKD equipment, we observe that in the CWDM backbone scenario, the limited absorption budget tolerable by the QKD systems can be as stringent as the  Raman scattering from classical channels that, apart from filtering, can be controlled through attenuation and very sensitive transponders without disrupting the classical transmission. The access scenario tested is essentially a worst case (for QKD) GPON in which no modification is done neither to the GPON management nor to the HW itself. In this case, a maximum 1:4 splitting is possible with the current QKD devices. In this scenario, it is possible to greatly increase the key throughput by modifying the device
profiles to assign time slots for QKD transmission. Without HW or SW modifications, the GPON scenario is mainly limited by Raman scattering and components crosstalk. In order to obtain the reported QBER and  key rates mentioned above, a decoy state protocol is needed. In such a heavily penalized scenarios, it is important not to waste bits of the raw key and more efficient error co\-rre\-ction protocols~\cite{cascade} would be a welcome addition~\cite{LDPCnuestros}. On the other hand, next generation QKD devices like the ones now in the labs, able to withstand more than 30 dB losses, would allow for far higher bit rates and would open the possibility of crossing a full metro network without the need of trusted repeaters. 

The authors are indebted to M. Curty for his helpful comments.
This work was supported by CDTI, Ministry of Trade and Industry of Spain under project Segur@, CENIT-2007 2004 and UPM 178/Q06 1005-127.

%
%

%

\begin{thebibliography}{5}


\bibitem{secoqc} A. Poppe, M. Peev and O. Maurhart: Outline of the SECOQC Quantum-Key-Distribution Network.  Int. J.  Quantum Inf. 6, No. 2, pp. 209--218 (2008)

\bibitem{toliver09} N.A. Peters {\sl et al.}: Dense Wavelength Multiplexing of 1550 nm QKD with Strong Classical Channels in Reconfigurable Networking Environments. New. J. of Phys. 11, 045012 (2009)

\bibitem{lancho08} D. Lancho {\sl et al.}: Quantum Key Distribution in Commercial Optical Networks. Report to the SECOQC Conference, Oct. 2008. Available at http://www.secoqc.net
 
\bibitem{BB89} C.H. Bennett and G. Brassard: The Dawn of a New Era for Quantum Cryptography: The experimental prototype is working!
Sig. Act News 20(4), p. 78 (1989) 

\bibitem{toliver03} P. Toliver {\sl et al.}: Experimental Investigation of Quantum Key Distribution Through Transparent 
Optical Switch Elements. IEEE Photonics Tech. Lett. 15, Is. 11, pp. 1669--1671 (2003)

\bibitem{runser06} R.J. Runser {\sl et al.}: Quantum Key Distribution for Reconfigurable Optical Networks. 2006 Optical Fiber Communication Conference. Contribution OFL1 (2006)

\bibitem{runser05} R.J. Runser {\sl et al.}: Demonstration of 1.3 $\mu$m Quantum Key Distribution Compatibility with 1.5 $\mu$m Metropolitan Wavelength Division Multiplexed Systems. 2005 Optical Fiber Communication Conference. 
                   Contribution OWI2 (2005)

\bibitem{subacius05} D. Subacius, A. Zavriyev and A. Trifonov: Backscattering Limitation for Fiber-optic Quantum Key Distribution Systems. Appl. Phys. Lett.  86, 011103 (2005)

\bibitem{xia05} T.J. Xia {\sl et al.}: In-Band Quantum Key Distribution (QKD) on Fiber Populated by High-Speed Classical Data Channels. 2006 Optical Fiber Communication Conference. Contribution OTuJ7 (2006) 

\bibitem{toliver07} P. Toliver {\sl et al.}: Demonstration of 1550 nm QKD with ROADM-based DWDM Networking and the Impact of Fiber FWM. 2007 Optical Fiber Communication Conference. Contribution CThBB1 (2007)

\bibitem{ITU-WDMStandards} DWDM: ITU-T G.694.1 (06/2002): Spectral grids for WDM applications: DWDM 
frequency grid. http://www.itu.int/rec/T-REC-G.694.1-200206-I/en

\bibitem{ITU_CWDM} CWDM: ITU-T G.694.2 (12/2003): Spectral grids for WDM applications: CWDM wavelength grid. http://www.itu.int/rec/T-REC-G.694.2-200312-I/en

\bibitem{GPONStandard} GPON: ITU-T G.984.1 (03/2008): Gigabit-capable Passive Optical Networks (GPON): General Characteristics. http://www.itu.int/rec/T-REC-G.984.1-200803-I/en

\bibitem{cascade} G. Brassard and L. Salvail: Secret-key reconciliation by
public discussion. In Eurocrypt93, Workshop on the theory and application
of cryptographic techniques on Advances in cryptology, ser. Lecture Notes
in Computer Science, vol. 765, 1994, pp. 410-423 (1994)

\bibitem{LDPCnuestros} D. Elkouss, J. Martinez, D. Lancho and V. Martin:
Rate Compatible Protocol for Information Reconciliation: An application to
QKD. Proceedings of the IEEE Information Theory Workshop 2010 (ITW 2010, Cairo), pp. 145-149. (2010)

\bibitem{BBBSS92} C.H. Bennett, F. Bessette, G. Brassard, L. Salvail and J.  Smolin: First QKD experiment. J. of Cryptology 5, p. 3 (1992) 

\bibitem{decoyLo} H-K. Lo, X. Ma, and K. Chen: Decoy State Quantum Key Distribution Phys. Rev. Lett. 94, 230504 (2005)

\bibitem{decoyHwang} W.-Y. Hwang: Quantum Key Distribution with High Loss: Toward Global Secure Communication, Phys. Rev. Lett., Vol. 91, No. 5 (2003)

\bibitem{decoyWang} X. B. Wang: Beating the photon-number-splitting attack in practical quantum cryptography, Phys. Rev. Lett. 94, 230503 (2005)

\bibitem{Wang05} X.B. Wang: Decoy-state protocol for quantum cryptography with four different intensities of coherent light, Phys. Rev. A 72, 012322 (2005)

\bibitem{Ma05} X. Ma, B. Qi, Y. Zhao and H-K Lo: Practical Decoy State for Quantum Key Distribution. Phys. Rev. A 72, 012326 (2005)

\bibitem{WegmanCarter} M. N. Wegman and J. L. Carter: New hash functions and
their use in authentication and set equality, Journal of Computer and System
Sciences, Vol. 22, pp. 265--279 (1981).

\bibitem{vanassche} G. Van Assche: Quantum Cryptography and Secret-Key Distillation. Cambridge University Press (2006)



\end{thebibliography}
\end{document}